\documentclass[%
twocolumn,
 amsmath,amssymb,
 aps,authoryear
]{aastex631}
\usepackage{graphicx}
\usepackage{ulem}
\usepackage{epstopdf}
\usepackage{amsmath}
\usepackage{amssymb}
\usepackage{mathtools}
\usepackage{mathrsfs}
\usepackage{bm}
\setcitestyle{authoryear,round} 
\usepackage{color}
\usepackage{xcolor}
\usepackage{textcomp}
\usepackage{gensymb}
\usepackage{hyperref}
\newcommand{\rs}{\text{R}_\odot}
\newcommand{\dash}{\text{-}}
\newcommand{\epar}{\hat{\bm{e}}_\parallel}
\newcommand{\eperpa}{\hat{\bm{e}}_{\perp 1}}
\newcommand{\eperpb}{\hat{\bm{e}}_{\perp 2}}
\newcommand{\bperpa}{b_{\perp 1}}
\newcommand{\bperpb}{b_{\perp 2}}
\begin{document}

\title{Anisotropic Magnetic Turbulence in the Inner Heliosphere -- Radial Evolution of Distributions observed by Parker Solar Probe}

\author[0000-0002-7174-6948]{Rohit Chhiber}
\affiliation{Department of Physics and Astronomy, University of Delaware, Newark, DE 19716, USA}
\affiliation{Heliophysics Science Division, NASA Goddard Space Flight Center, Greenbelt, MD 20771, USA}
\email{rohit.chhiber@nasa.gov}

%
\begin{abstract}
Observations from Parker Solar Probe's first five orbits are used to investigate the helioradial evolution of probability density functions (PDFs) of fluctuations of magnetic field components, between \(\sim 28\) - 200 \(\rs\). Transformation of the magnetic field vector to a local mean-field coordinate system permits examination of anisotropy relative to the mean magnetic field direction. Attention is given to effects of averaging-interval size. It is found that PDFs of the perpendicular fluctuations are well approximated by a Gaussian function, with the parallel fluctuations less so: kurtoses of the latter are generally larger than 10, and their PDFs indicate increasing skewness with decreasing distance \(r\) from the Sun, \added{with the latter observation possibly explained by the increasing Alfv\'enicity of the fluctuations}. The ratio of perpendicular to parallel variances is greater than unity; this variance anisotropy becomes stronger with decreasing \(r\). The ratio of the total rms fluctuation strength to the mean field magnitude decreases with decreasing \(r\), with a value \(\sim 0.8\) near 1 AU and \(\sim 0.5\) at 0.14 AU; the ratio is well approximated by a \(r^{1/4}\) power law. These findings improve our understanding of the radial evolution of turbulence in the solar wind, and have implications for related phenomena such as energetic-particle transport in the inner heliosphere. 
\end{abstract}


\section{Introduction}

The probability density function (PDF) occupies a fundamental role in statistical treatments of turbulence \citep{Monin1971book}. In hydrodynamics, one is usually concerned with PDFs of the turbulent velocity. In the solar wind, the magnetic field, whose measurements are more readily available, often takes on the role of the primitive fluctuating field, to the extent that the magnetohydrodynamic (MHD) description is valid \citep{goldstein1995araa}. A Gaussian distribution is a common reference, and departures from Gaussianity indicate the strength of non-linearity \citep{zhou2004RMP}. The degree of non-Gaussianity is also related to theoretical approaches such as the ``quasi-normal'' hypothesis, wherein fourth-order moments are assumed to be Gaussian, while third-order moments can depart from the Gaussian value of zero, giving rise to various turbulence closure models \citep{lesieur2008book}. In recent years, stochastic approaches have been developed to study energetic particle transport within realizations of random magnetic fluctuations, which are often assumed be Gaussian \citep[e.g.,][]{tooprakai2016ApJ}. For these reasons, it is important to establish a firm observational basis for the PDF of magnetic fluctuations in the solar wind. 

One feature of solar wind turbulence that distinguishes it from hydrodynamic turbulence is the \textit{anisotropy} introduced by a mean magnetic field \citep[e.g.,][]{oughton2015philtran}. This anisotropy  can be of two types -- \textit{spectral} anisotropy refers to unequal distribution of power in spectra of a turbulent field when examined as functions of wavenumbers parallel and perpendicular to the magnetic field. This is a consequence of the anisotropic transfer of energy in the MHD turbulent cascade, \added{and is often associated with anisotropy in the correlation function \citep[e.g.,][]{matthaeus1990JGR,dasso2005ApJ}.} \textit{Variance} (or \textit{component}) anisotropy refers to unequal energies in the components of a fluctuating field, and can be related to the relative dominance of various fluctuating wave modes; for instance, in the solar wind one usually finds stronger fluctuations transverse to the mean magnetic field \citep{belcher1971JGR}, \deleted{indicating predominantly Alfv\'enic modes and weak compressibility}  \added{which are associated with weak compressibility and either Alfv\'en waves or two-dimensional turbulence \citep{oughton2015philtran}}. By analysing PDFs of the magnetic field in a coordinate system defined by the  mean magnetic field, it becomes possible to obtain insights regarding variance anisotropy. Note that variance anisotropy does not necessarily imply spectral anisotropy \citep[see][]{oughton2015philtran}.

Several previous works have used near-Earth observations to investigate PDFs of the magnetic field and the related anisotropy \citep{whang1977SoPh,feynman1994JGR,padhye2001JGR}. In the Parker Solar Probe (PSP) era it has become possible to extend these studies much closer to the Sun \citep{fox2016SSR}. The goal of this paper is to use PSP measurements of the magnetic field, accumulated over its first five orbits, to investigate PDFs of magnetic fluctuations in the inner heliosphere and to trace their radial evolution between \(\sim 28\) - \(200~\rs\) (\(\sim 0.14~ \dash~ 0.9\) AU). Our focus will be on PDFs of the primitive fluctuations and not their \textit{increments}; the latter are commonly used to study small-scale intermittency \citep[e.g.,][]{matthaeus2015ptrs}. Further, we restrict ourselves to a study of component anisotropy. PSP measurements near the Sun have been used to study spectral (correlation) anisotropy as well \citep[e.g.,][]{Bandyopadhyay2021ApJ,cuesta2022ApJL,adhikari2022ApJ_aniso}.

In outline, we briefly describe the data set employed in Section \ref{sec:data} and discuss the transformation to mean-field coordinates in Section \ref{sec:mfc}. The remainder of Section \ref{sec:results} presents our investigation of the radial evolution of PDFs of the magnetic field and its first four statistical moments. We conclude with discussion in Section \ref{sec:disc}.




\section{Parker Solar Probe Data}\label{sec:data}
We used publicly available magnetic field data from the fluxgate magnetometer (MAG), part of the FIELDS instrument suite aboard PSP \citep{bale2016SSR}, for the first five orbits covering the period between October 2018 to July 2020. The perihelia were at \(35.6\ \rs\) for orbits 1 to 3, and at \(\sim 28\ \rs\) for orbits 4 and 5. The spacecraft stayed close to the ecliptic plane in its highly elliptical orbit \citep{fox2016SSR}. Level 2 MAG data were resampled to 1-s cadence using a linear interpolation. Missing or bad data were represented as NaNs. In the following analyses, data from the five orbits were aggregated and statistical computations were performed within radial bins of size \(5~\rs\) or \(10~\rs\), as indicated below. Except for three (\(5~\rs\)) bins located between \(150~\dash~ 170~\rs\), fewer than 50\% of data within each bin were NaNs. All bins at heliocentric distances smaller than \(70~\rs\) had less than 10\% of data as NaN.

\section{Results}\label{sec:results}

\subsection{Transformation to local mean magnetic-field coordinates}\label{sec:mfc}

The local mean magnetic field is computed as \(\bm{B}_0 \equiv \langle \bm{B}\rangle\), where \(\langle\cdot\rangle\) refers to a moving boxcar average over a 2-hour window centered at each respective instant in the time series. This interval size is sufficiently longer than the correlation scale of the turbulence \citep{chen2020ApJS,cuesta2022ApJL}, while also much shorter than the solar rotation period \citep[see also][]{padhye2001JGR}. The averaging produces a time series of the (local) mean magnetic field \(\bm{B}_0\) at 1-second cadence. We then specify a local mean-field coordinate (MFC) system at \textit{each point} in the time series, using a standard approach \citep{belcher1971JGR,bruno2013LRSP}: The unit vector parallel to \(\bm{B}_0\) is \(\epar \equiv (B_{0,R} \hat{\bm{e}}_R + B_{0,T} \hat{\bm{e}}_T + B_{0,N} \hat{\bm{e}}_N)/B_0\), where \(\{R,T,N\}\) refers to heliocentric \(RTN\) coordinates \citep{franz2002pss}, and \(B_0\) is the magnitude of the mean field. We define the ``second'' perpendicular direction as \(\eperpb\equiv \epar \times \hat{\bm{e}}_R/|\epar \times \hat{\bm{e}}_R|\). The ``first'' perpendicular direction is then  \(\eperpa\equiv \eperpb \times\epar/|\eperpb \times\epar|\).

\begin{figure}
    \centering
       \includegraphics[width=\columnwidth]{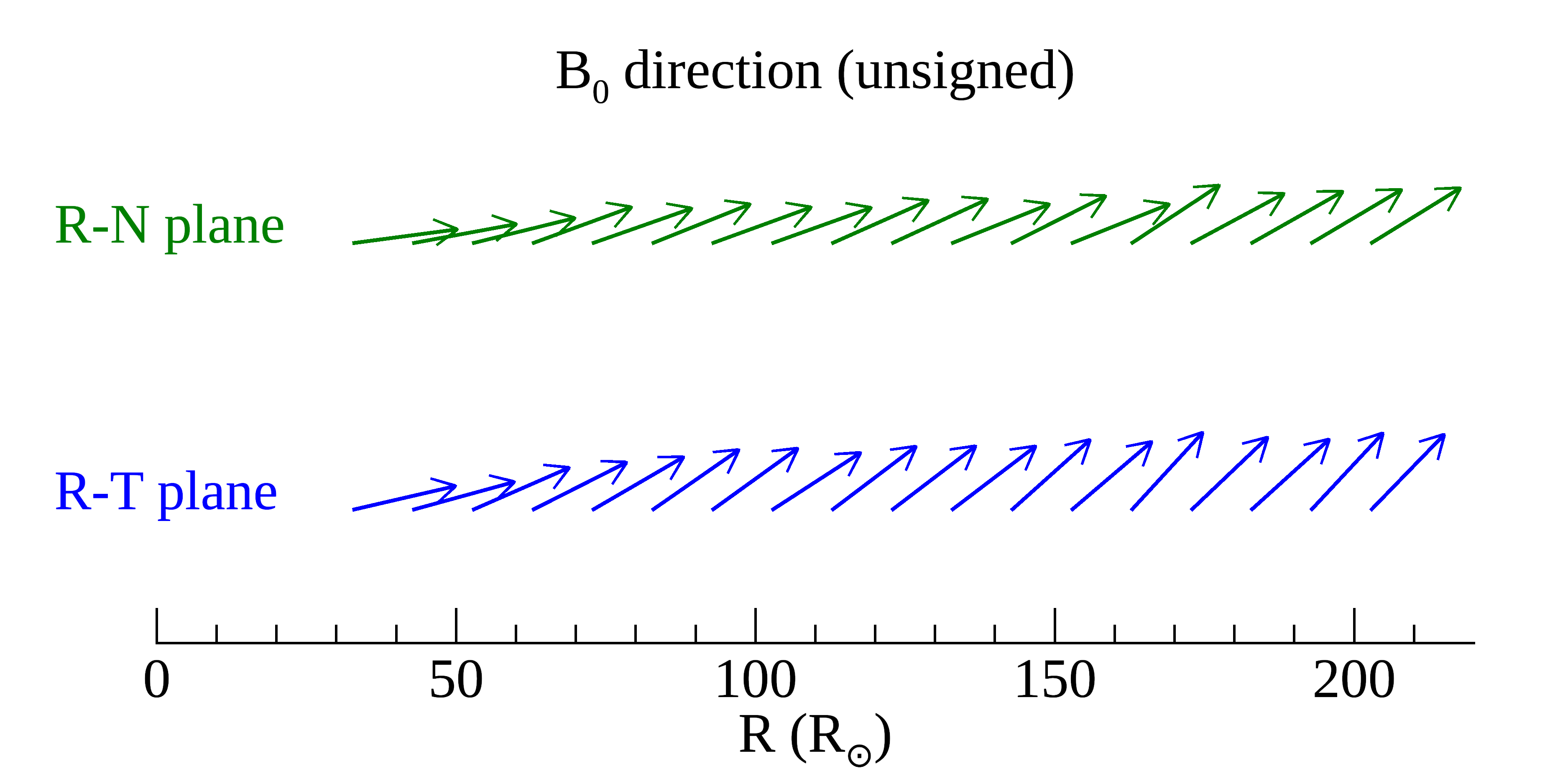}
    \caption{Direction of mean magnetic field \(\bm{B}_0\) as a function of radial distance, aggregated from PSP orbits 1 to 5, and averaged in radial bins of size \(10~\rs\). Horizontal axis represents \(R\) coordinate, and vertical direction represents \(T\) and \(N\) coordinates for blue and green cases, respectively. The sign of components of \(\bm{B}_0\) is not taken into account, and length of arrows is arbitrary. Tail of each arrow is placed at center of the respective radial bin. At \(30~\rs\), angles made by \(\bm{B}_0\) wrt to the \(R\) direction are 13\degree\ and 8\degree\ in the \(RT\) and \(RN\) planes, respectively,  while at \(200~\rs\) these angles are 46\degree\ and 32\degree.}
    \label{fig:B0}
\end{figure}

To provide global context regarding the orientation of the mean magnetic field, Figure \ref{fig:B0} shows the direction of \(\bm{B}_0\) projected on the \(R\dash T\) and \(R\dash N\) planes. Here \(\bm{B}_0\) is aggregated from all five PSP orbits, and each of its (unsigned) components are averaged in \(10~\rs\) bins of heliocentric distance.\footnote{This representation of the direction of \(\bm{B}_0\) does not distinguish between positive and negative polarities. Further, recall that \(\bm{B}_0\) is a 2-hour averaged quantity, and therefore the so-called switchbacks, which occur at smaller scales \citep{DudokDeWit2020ApJS}, are not present in it. \added{Switchbacks are contained in the full magnetic field expressed in Equation \eqref{eq:B_mfc}, however.}} One can visualize the formation of the Parker spiral \citep[e.g.,][]{owens2013lrsp} via this figure, as the field changes from mostly-radial at small helioradii, to displacements from the \(\hat{\bm{e}}_R\) direction of about 45\degree\ and 30\degree\ near Earth, in the \(R\dash T\) and \(R\dash N\) planes, respectively.\footnote{ \cite{cuesta2022ApJL} show the radial evolution of distributions of the angle between the mean magnetic field and the radial direction, as observed by PSP.}

Once the MFC system is established, we rotate the magnetic field into this system \citep{bruno2013LRSP}, which allows us to write 
\begin{equation}
    \bm{B}= B_\parallel\epar + \bperpa \eperpa + \bperpb \eperpb. \label{eq:B_mfc}
\end{equation}
\added{Since the two transverse components \(\bperpa\) and \(\bperpb\) are always perpendicular to the mean field, they are, by definition, the transverse fluctuations. The parallel fluctuation is computed as \(b_\parallel = B_\parallel - B_0 \). This procedure produces a timeseries of the fluctuations at 1-sec cadence.} 
\deleted{where \(b_\parallel\), \(\bperpa\), and \(\bperpb\) are the orthogonal fluctuations.} Note that since the mean field is nearly radial at small helioradii (\(\lesssim 60~\rs\)), \(\bperpb\) and \(\bperpa\) lie approximately in the \(\pm N\) and \(\pm T\) directions, respectively, at these distances. 

We remark here that the results obtained (described in subsequent sections) remain essentially unchanged on varying the averaging interval for computation of the mean field to 4 hours or 30 minutes, with one notable exception, namely, the ratio of rms fluctuation to mean magnetic field, discussed in detail below. Note also that analysis of fluctuation anisotropy may be performed in the \textit{minimum variance} coordinate frame instead of the MFCs; however, the angle between the minimum-variance direction and the mean-field direction has been shown to be small \citep{bavassano1982SoPh}.

\subsection{Radial evolution of statistical moments of magnetic fluctuations}\label{sec:moments}

The PSP measurements considered in this study cover heliocentric distances from \(\sim 28~\text{-} ~200 ~\rs\), over five complete orbits. Aggregating data accumulated over five orbits enables investigation of long-term radial trends \citep[see also][]{chhiber2021ApJ_psp}. Note that the measurements were made during solar-minimum conditions near the ecliptic plane, and therefore the data set is overwhelmingly representative \citep[\(>98\)\%; see][]{chhiber2021ApJ_psp} of slow wind conditions.

We first examine broad radial trends seen in the first four statistical moments of the magnetic fluctuations. The aggregated magnetic field data (at 1-s resolution) from the first five orbits are grouped within \(5~\rs\) bins, and the following quantities are computed for each fluctuation component (say, \(x\)), within each radial bin:
\begin{equation}
    \text{Mean} = \bar{x} = \frac{1}{N} \sum_{j=1}^{j=N} x_j,
\end{equation}

\begin{equation}
    \text{Std Dev} = \sqrt{\frac{1}{N} \sum_{j=1}^{j=N} (x_j - \bar{x})^2},
\end{equation}

\begin{equation}
    \text{Skewness} = \frac{1}{N} \sum_{j=1}^{j=N} 
       \left( \frac{x_j - \bar{x}}{\text{Std Dev}}\right)^3,
\end{equation}
and
\begin{equation}
    \text{Kurtosis} = \frac{1}{N} \sum_{j=1}^{j=N} 
       \left( \frac{x_j - \bar{x}}{\text{Std Dev}}\right)^4,
\end{equation}
where \(N\) is the number of data points within each bin. Results shown below were found to not vary significantly on changing the bin size to \(0.5~\rs\) and \(20~\rs\). Here `std dev' refers to the standard deviation, and its square is the variance. 

\begin{figure}
    \centering
    \includegraphics[width=\columnwidth]{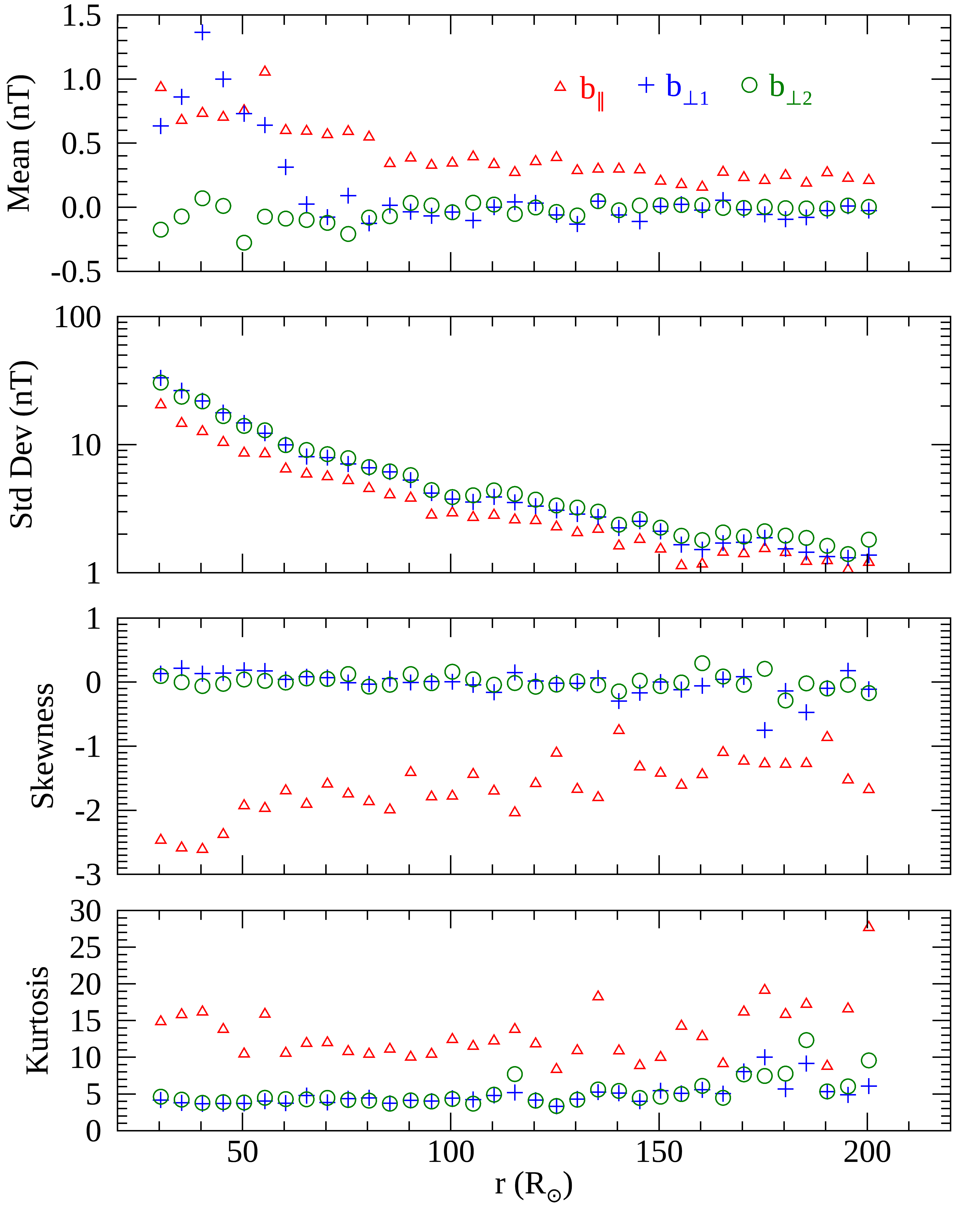}
    \caption{Statistical moments of magnetic fluctuation components, computed in radial bins of size \(5\ \rs\) by aggregating data from PSP's first five orbits. Here `std dev' the standard deviation. Error bars (not shown) are generally smaller than the symbols.}
    \label{fig:moments}
\end{figure}

The top panel of Figure \ref{fig:moments} shows that the mean value of the fluctuations remains small for all distances shown, relative to the mean field magnitude observed by PSP, which increases from \(\sim 5\) nT near 1 AU to \(\sim 60\) nT at \(30~\rs\) \citep[see, e.g.,][]{chhiber2021ApJ_psp}. \added{Recall that the local MFC system used here implies that each computed pointwise fluctuating field value is associated with a different mean field. Therefore, when one computes the mean fluctuation by aggregating the pointwise fluctuations within a particular \(5 ~R_\odot\) bin, the result is not exactly zero. This reflects the variability of the mean field within a radial bin.} Note that the components exhibit differences: \(b_\parallel\) has a small, non-zero mean, which increases gradually with decreasing helioradius \(r\), while \(\bar{b}_{\perp 1}\) begins to increase below \(\sim 60~\rs\). Since the latter component lies approximately in the \(\pm T\) direction close to the Sun, the larger mean may be related to strong azimuthal flows found in that region \citep{weber1967ApJ148,kasper2019Nat}. 

\begin{figure}
    \centering
    \includegraphics[width=\columnwidth]{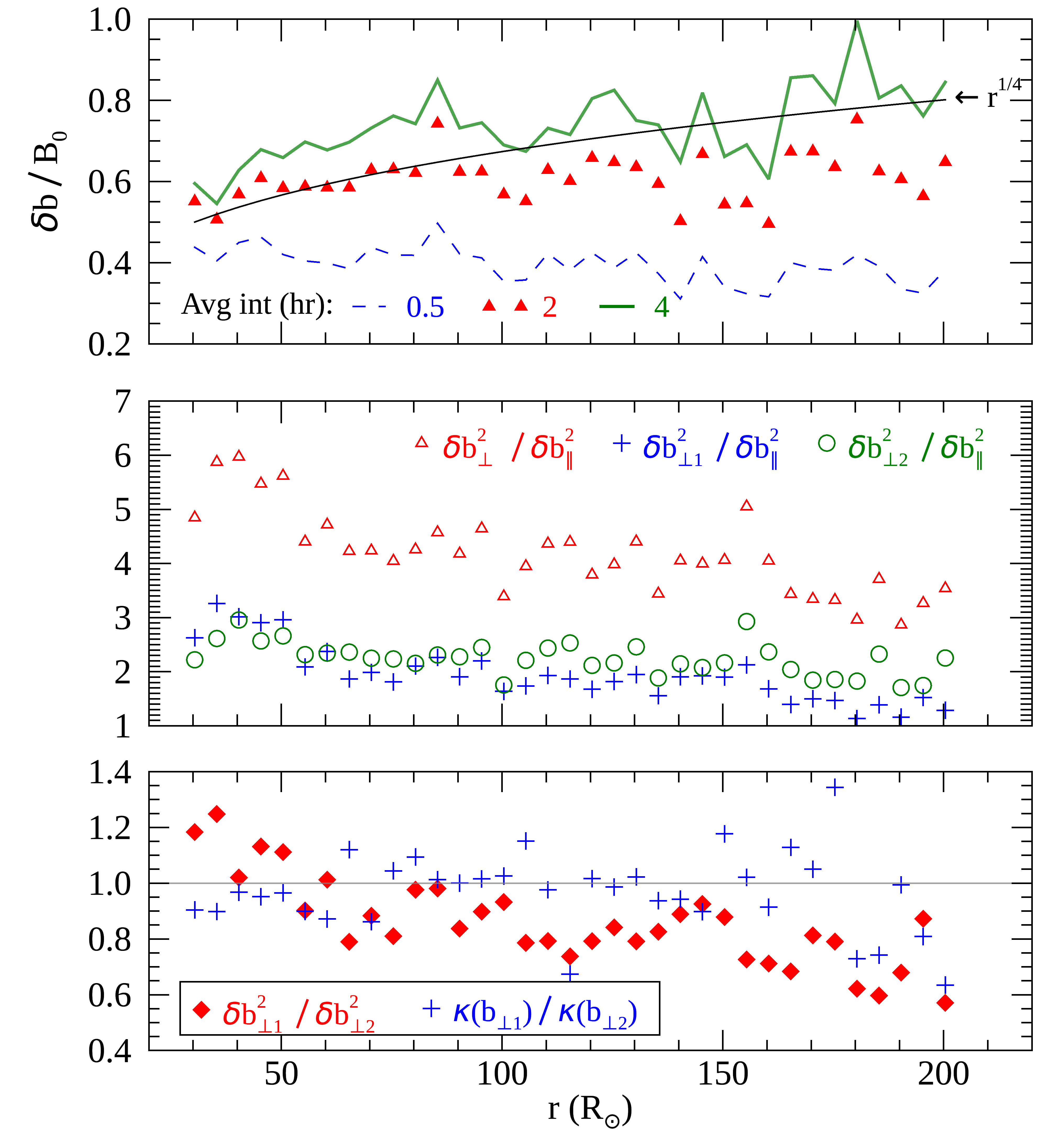}
    \caption{\textit{Top}: Ratio of rms magnetic fluctuation to mean magnitude of the magnetic field. Three cases are shown, based on different temporal averaging intervals for computation of mean and fluctuating fields. Black curve shows \(r^{1/4}\) power-law scaling (see text). \textit{Middle}: Measures of variance anisotropy. \textit{Bottom}: Test of non-axisymmetry of transverse fluctuations. Here \(\kappa\) refers to the kurtosis. See text for other definitions. In all panels, statistics are computed within radial bins of size \(5~\rs\); error bars (not shown) are smaller than symbols.}
    \label{fig:axisym}
\end{figure}

The second panel from the top of Figure \ref{fig:moments} shows the standard deviation of the fluctuations, which increases nearly monotonically as one approaches the Sun, indicating an order-of-magnitude increase in fluctuation strength between 1 AU and 0.14 AU. A systematic variance anisotropy is observed, and is investigated in greater detail in Figure \ref{fig:axisym}.

The third panel from the top in Figure \ref{fig:moments} shows the radial evolution of skewness, which is a measure of the asymmetry of a distribution \citep[e.g.,][]{doane2011JSE_skew} and is related to the triple correlations that arise from non-linearities in turbulence theory \citep[e.g.,][]{zhou2004RMP}. A striking difference is observed between parallel and transverse fluctuations: the former possess a (signed) skewness smaller than \(-1\) at \(200~\rs\), which systematically decreases to about \(-2.6\) at \(30~\rs\). This trend may be \added{a consequence of the Alfv\'enic nature of fluctuations observed by PSP near the Sun (see discussion in Section \ref{sec:disc}), and possibly} related to the abrupt magnetic-field reversals (``switchbacks'') ubiquitous in PSP observations \citep{DudokDeWit2020ApJS}. In contrast, both transverse components maintain a near-zero skewness for all distances shown. Note that the larger skewness in \(b_\parallel\) is apparent in Omnitape and Ulysses observations analyzed by \cite{padhye2001JGR} (see Figures 1, 4, and 6 of that paper), although the authors do not comment on this finding. 

Finally, the bottom panel of Figure \ref{fig:moments} shows the kurtosis of the fluctuations; a Gaussian distribution has a kurtosis of 3, while larger values indicate wider tails (intermittency) and peakedness in the distribution, relative to the Gaussian \cite[e.g.,][]{decarlo1997kurtosis}. There is some scatter in the values observed above \(170~\rs\), but for most distances, the kurtosis of the transverse fluctuations stays close to the Gaussian value. The kurtosis of \(b_\parallel\) is systematically higher (between \(10~\dash~ 20\)). This finding is consistent with previous studies \citep{marsch1994AnGeo,padhye2001JGR,bruno2003JGR}, and may be explained by the fact that transverse fluctuations could be associated with stochastic (non-intermittent) Alfv\'en waves \citep[see][]{bruno2003JGR}.

In Figure \ref{fig:axisym} we move on to an examination of the strength of magnetic fluctuations relative to the mean field, and of various measures of fluctuation anisotropy and axisymmetry. For convenience, we introduce the following notation in the figure: variances \added{(computed simply as the square of the standard deviations shown in Figure \ref{fig:moments})} of the respective fluctuation components are denoted as \(\delta b_\parallel^2,~\delta \bperpa^2,~\text{and}~\delta\bperpb^2\), so that the total rms magnetic fluctuation is \(\delta b \equiv (\delta b_\parallel^2 + \delta \bperpa^2 + \delta\bperpb^2)^{1/2}\), while the total transverse variance is \(\delta b_\perp^2\equiv \delta \bperpa^2 + \delta \bperpb^2\). The mean magnitude of the (mean) magnetic field within each radial bin is computed by first computing the magnitude \(|\bm{B}_0|\) of the mean field (see Section \ref{sec:mfc}) at each point within the bin, and then computing the mean of these pointwise values to obtain \(B_0\) for the bin. 

We first consider radial evolution of \(\delta b/B_0\). While all other results shown in this work remain essentially unchanged on varying the \added{duration of the temporal averaging window} \deleted{averaging interval} (say, \(\mathcal{T}\)) used for computation of the mean and fluctuating fields (Section \ref{sec:mfc}), this is not the case for \(\delta b/B_0\). This is not unexpected, since \(\delta b\) and \(B_0\) respond in opposite ways to changes in \(\mathcal{T}\), as discussed by \cite{isaacs2015JGR120}: longer intervals result in decreasing \(B_0\), since variations in the direction of the magnetic field vector \(\bm{B}\) ``cancel out'' when the mean field vector \(\bm{B}_0\equiv \langle \bm{B}\rangle\) is computed. Conversely, increasing \(\mathcal{T}\) implies larger fluctuation amplitude; intuitively, there is an ``exchange'' between \(\delta b\) and \(B_0\). This effect is evident in Figure \ref{fig:axisym}, where, at any given \(r\), longer \(\mathcal{T}\) results in larger \(\delta b/B_0\). Further, as one approaches the Sun, the ratio shows an increasing trend for \(\mathcal{T}=0.5\) hr, a \added{very slight decreasing trend} \deleted{nearly constant} for \(\mathcal{T}=2\) hr, and a decreasing trend for \(\mathcal{T}=4\) hr. We can get better insight into the radial evolution of \(\delta b/B_0\) by recalling that the averaging interval should ideally be several times larger than a correlation time \citep{matthaeus1982JGR}, which decreases from roughly 30-60 minutes at \(200~\rs\) to about 5-10 minutes at \(30~\rs\) \citep{chen2020ApJS,cuesta2022ApJL}. Therefore, one should consider the 4-hr case at larger distances (say, \(r \gtrsim 150~\rs\)), the 2-hr case for intermediate \(40~\rs \lesssim r \lesssim 150~\rs\), and the 0.5-hr case for \(r \lesssim 40~\rs\).\footnote{We acknowledge the unavoidable ambiguity in choosing these boundaries in \(r\).} All three cases appear to converge at small \(r\); future perihelia at smaller \(r\) may reveal further insights. We can conclude from the present analysis that \(\delta b/B_0\) decreases from \(\sim 0.8\) to \(\sim 0.5\) between \(200~\rs\) and \(30~\rs\): the relative fluctuation amplitude is strong near 1 AU, and remains moderately large down to the smallest distance considered.

Various models of turbulence transport predict different power-law scalings with \(r\) for the magnetic variance \(\delta b^2\) \citep[e.g.,][]{zhou1990JGR,zank1996evolution}. By assuming a radial scaling for the mean magnetic field, it is possible to compare the observed radial dependence of \(\delta b/B_0\) with theoretical predictions. For a brief and preliminary investigation, we assume that the mean field is radial \citep[a good assumption near the Sun, also used near 1 au for simplicity; e.g.,][]{tooprakai2016ApJ}, with a Parker-spiral type \(1/r^2\) scaling \citep[e.g.,][]{owens2013lrsp}. We can then write \(\delta b/B_0 \propto r^{2-\alpha/2}\), where \(\alpha\) defines the power-law scaling \(\delta b^2\propto 1/r^\alpha\) and can take on different values depending on the turbulence transport model being considered. Here we use asymptotic values of \(\alpha\) given by  \cite{zank1996evolution}\footnote{Values of \(\alpha\) used here require assumption of a \(1/r^2\) power law for density. Further, we only consider undriven models, since shear-driving introduces additional free parameters, and pickup-ion driving is not relevant in the inner heliosphere \citep[see][]{zank1996evolution}.}. We specify \(\delta b/B_0 = 0.5\) at \(r=30 ~\rs\) as a reference value. For the WKB model without turbulent dissipation or mixing between inward and outward Elsasser modes \citep[see][]{zank1996evolution}, \(\alpha = 3\), which produces an excessively large increase in \(\delta b/B_0\) with distance, compared to the observations, resulting in a value greater than 1.2 at 1 AU (not shown). When dissipation is present without mixing, \(\alpha = 3.5\), resulting in \(\delta b/B_0\propto r^{1/4}\) as plotted in Figure \ref{fig:axisym}, which agrees remarkably well with the observations. Finally, when mixing and dissipation are both present, \(\alpha = 4\), which produces a constant \(\delta b/B_0\), once again in disagreement with observations. We end with two caveats, first reminding the reader that a purely radial mean field was assumed for the present discussion, when, in fact, the tangential component can be significant at 1 AU. \added{Second, the model power-law scalings considered here were derived based on the assumption of zero cross helicity \citep{zank1996evolution}, which is not a good approximation near the Sun. Models that do not neglect cross helicity yield more complicated power-law scalings \citep[e.g.,][]{zhou1990JGR}.} Nevertheless, without reference to any particular model, the \textit{empirical} finding that a \(\delta b/B_0\propto r^{1/4}\) scaling agrees well with observations remains unaffected by these caveats.

We proceed to discuss the middle panel of Figure \ref{fig:axisym}, which examines variance anisotropy \citep[e.g.,][]{oughton2015philtran}. The ratios \(\delta \bperpa^2/\delta b_\parallel^2\) and \(\delta \bperpb^2/\delta b_\parallel^2\) are shown separately, and both indicate a modest but systematic increase as one approaches the Sun. The ratio \(\delta b_\perp^2/\delta b_\parallel^2\) increases from \(\sim 3\) to \(\sim 6\), signifying that the (MHD inertial range) turbulence becomes increasingly anisotropic in the low plasma-\(\beta\) environment of the solar corona, as expected from theory and models \citep{zank1993nearly,chhiber2019psp2,zank2021PoP}. 

\begin{figure*}
    \centering
    \includegraphics[width=.32\textwidth]{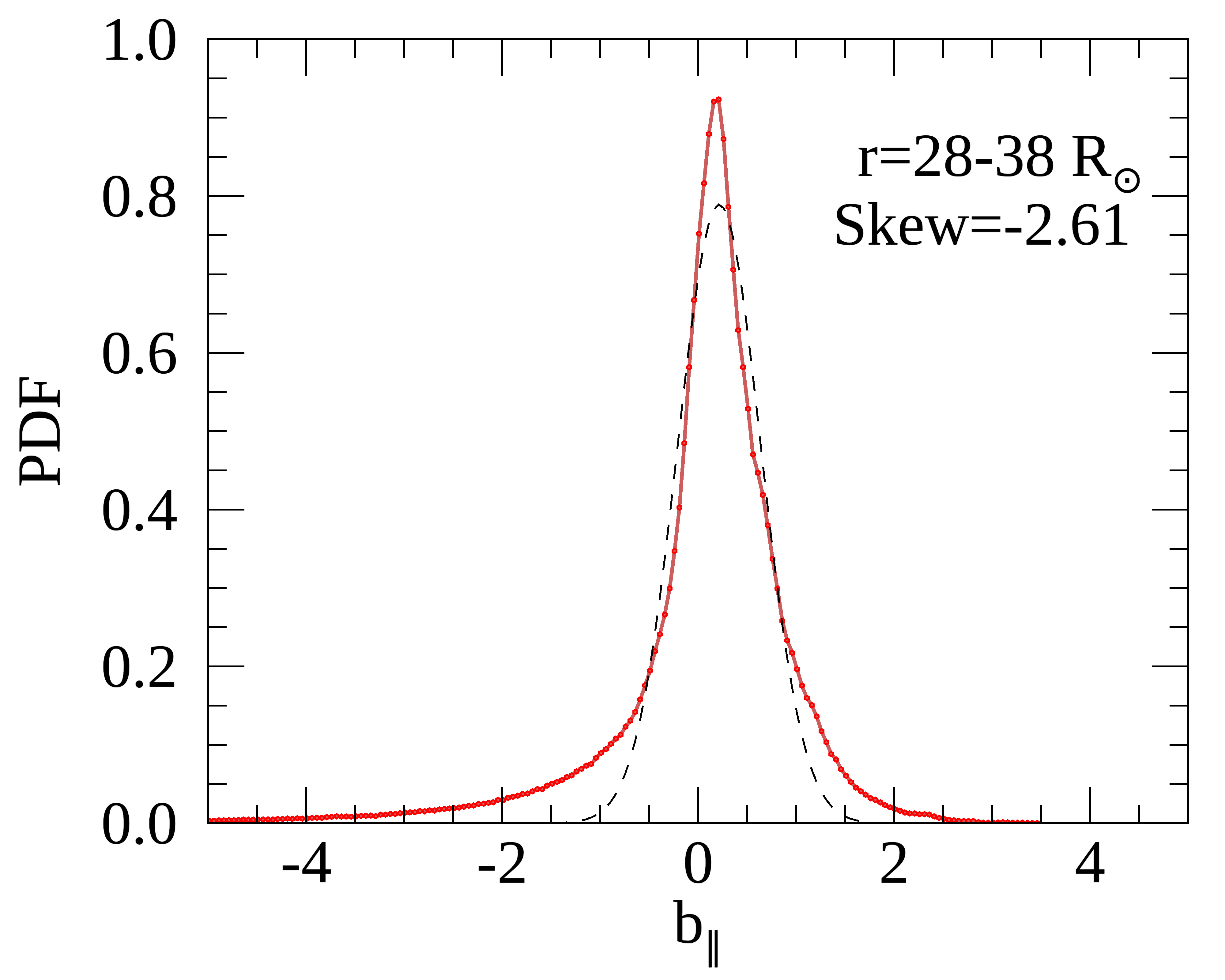}
    \includegraphics[width=.32\textwidth]{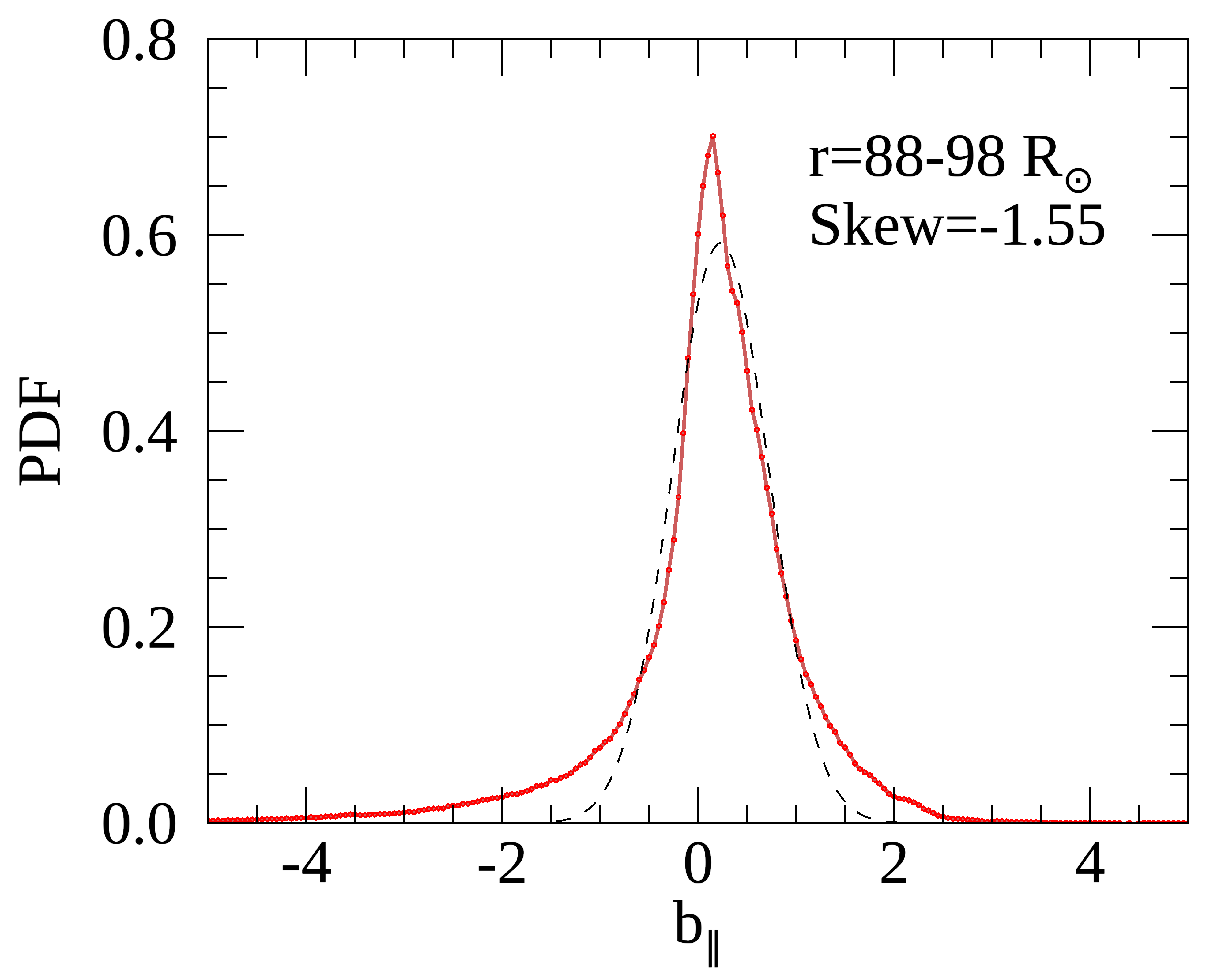}
    \includegraphics[width=.32\textwidth]{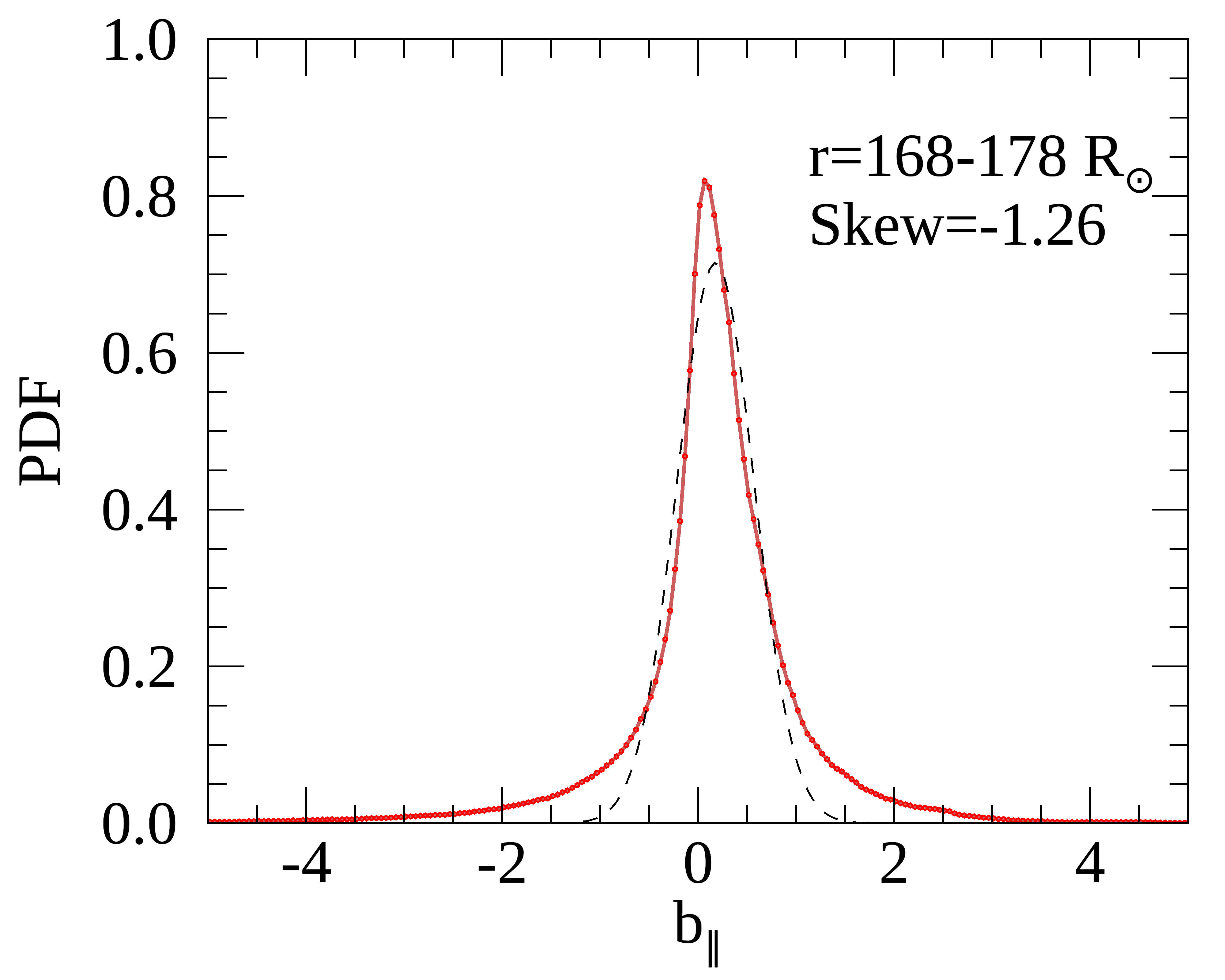}
    \includegraphics[width=.32\textwidth]{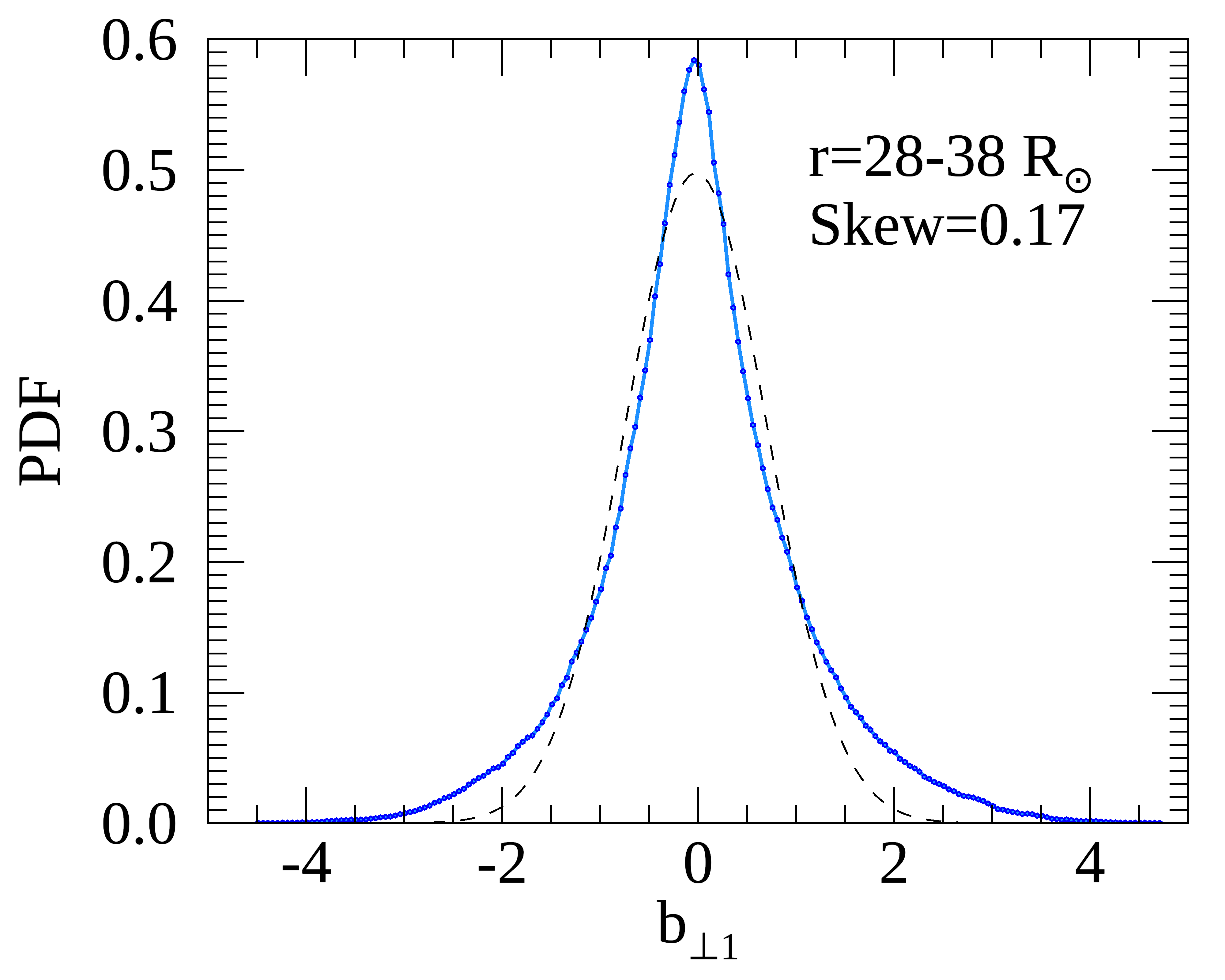}
    \includegraphics[width=.32\textwidth]{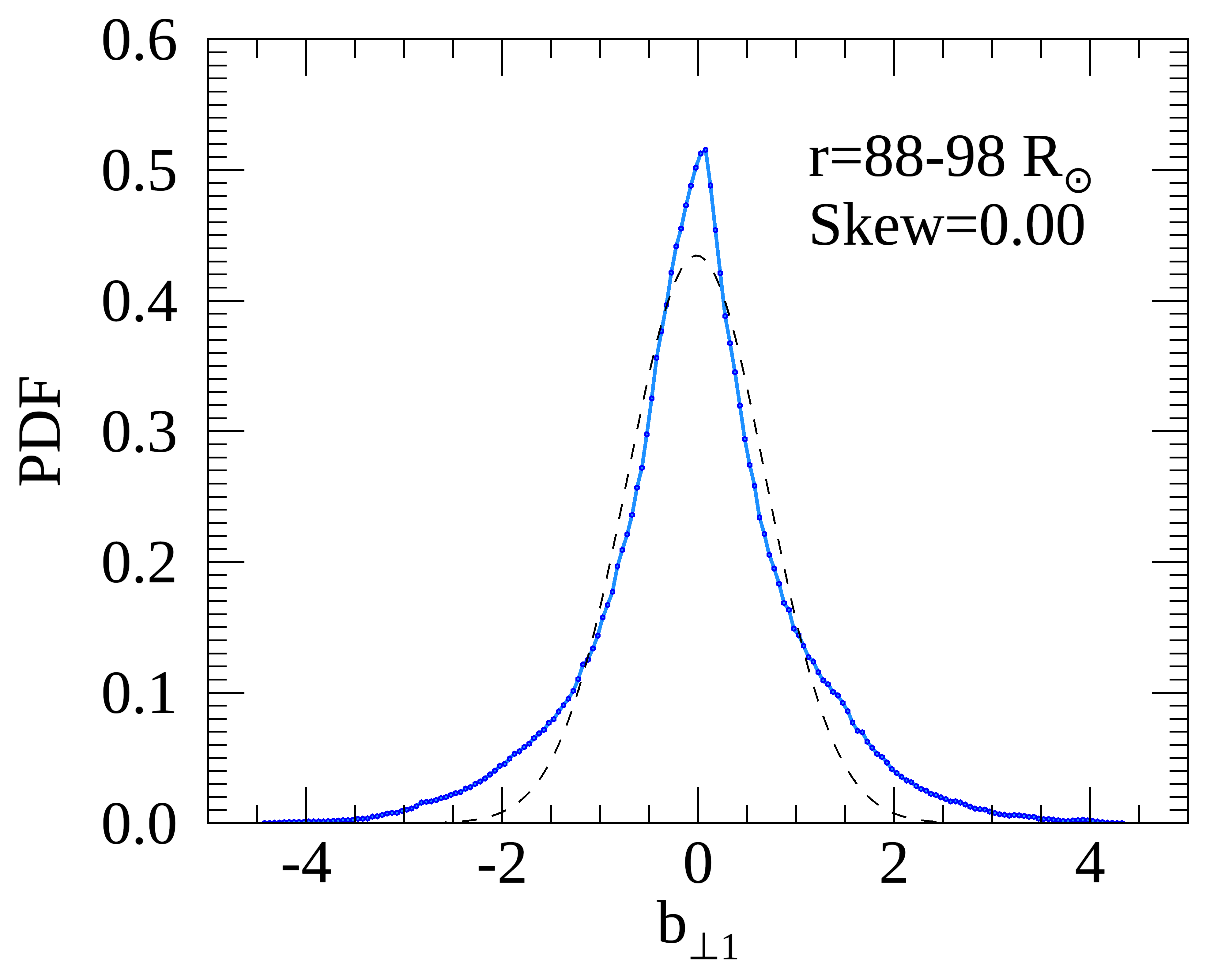}
    \includegraphics[width=.32\textwidth]{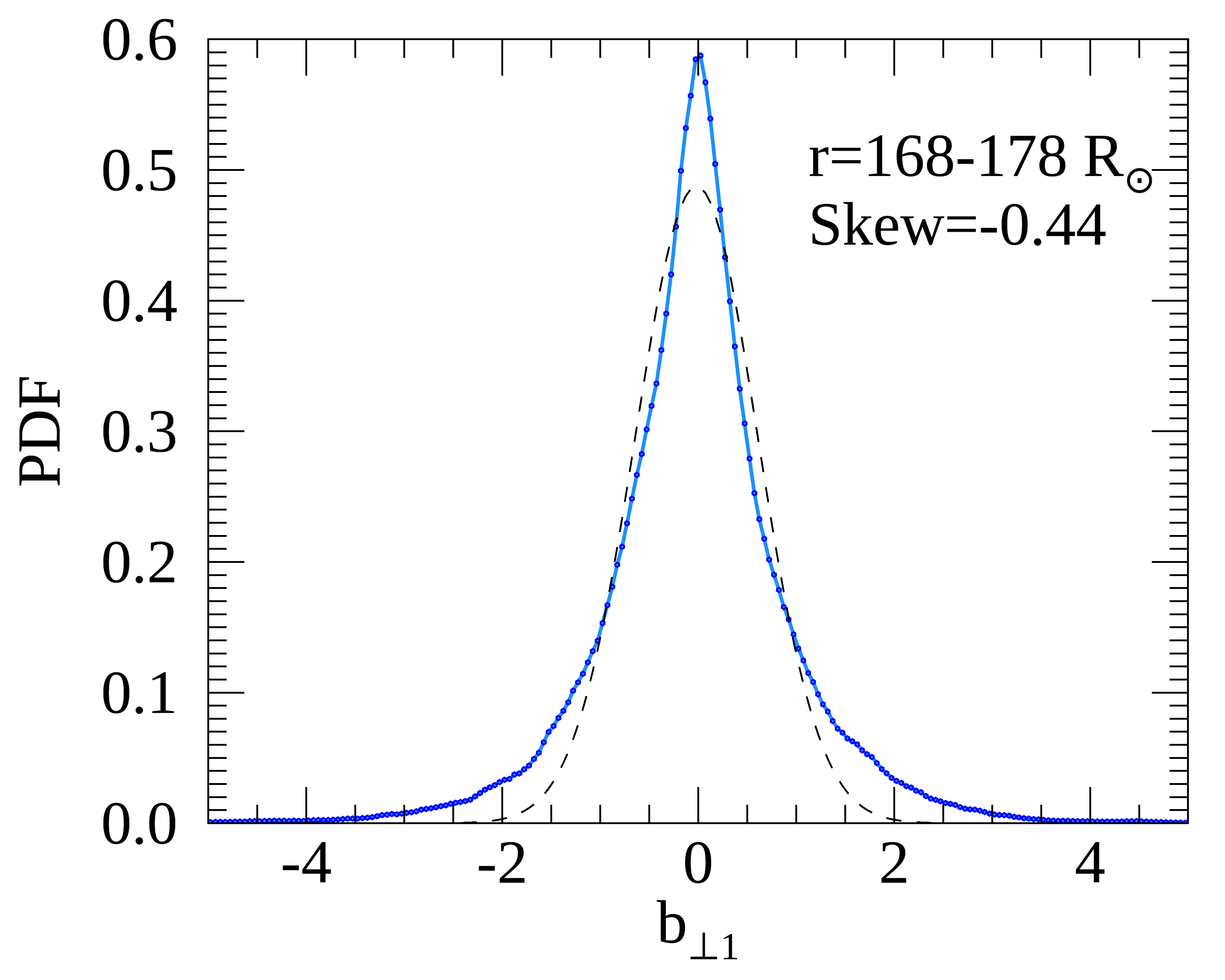}
    \includegraphics[width=.32\textwidth]{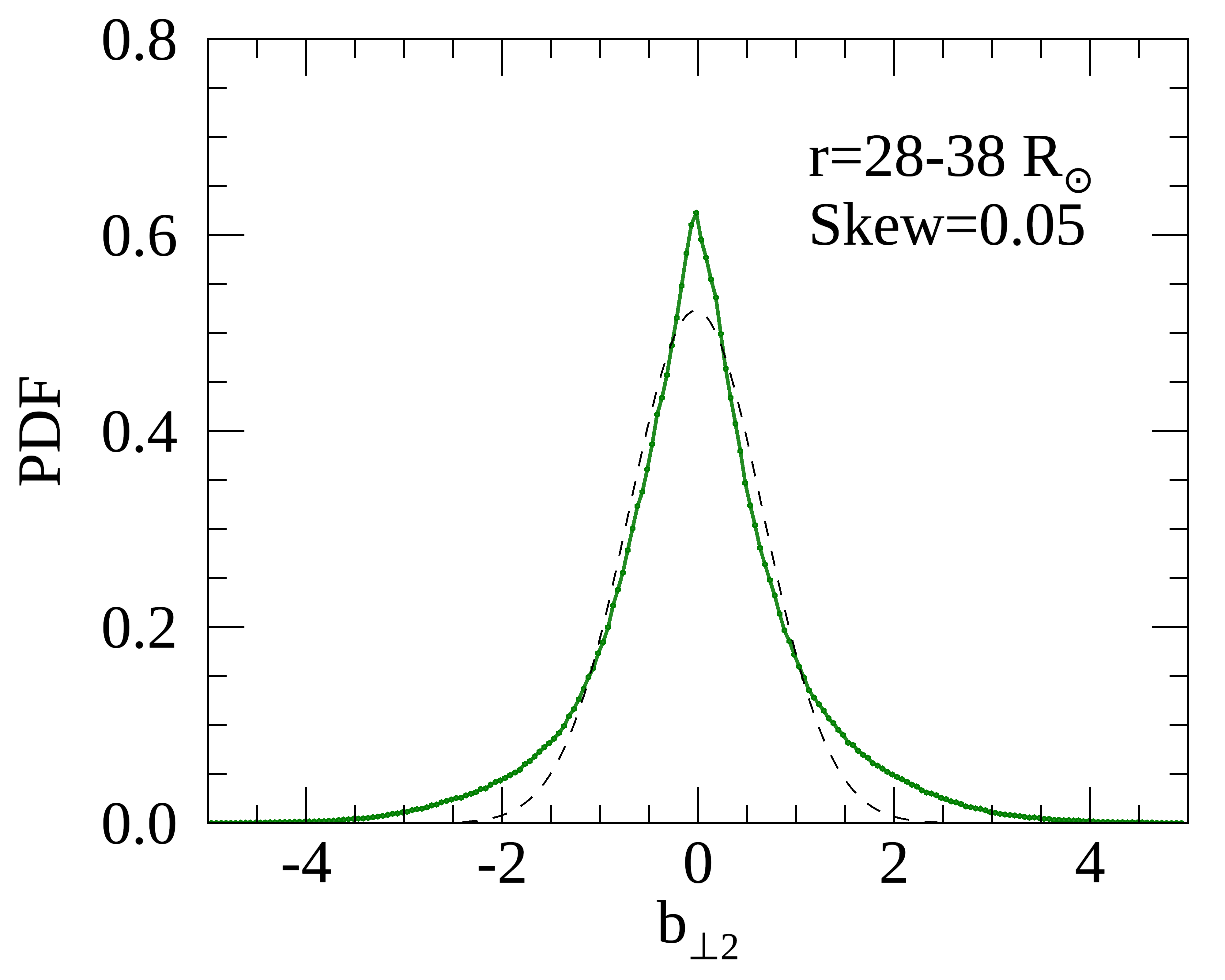}
    \includegraphics[width=.32\textwidth]{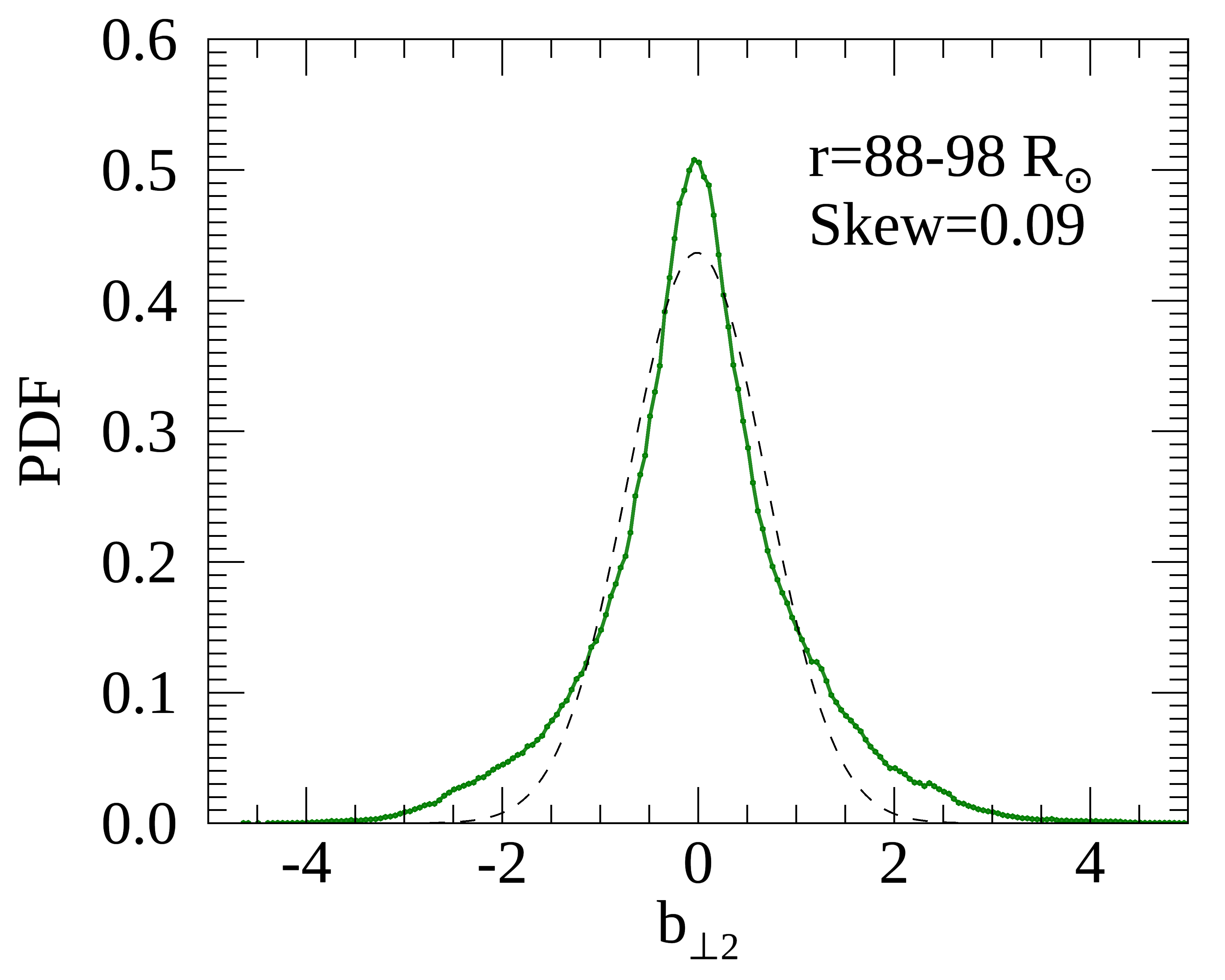}
    \includegraphics[width=.32\textwidth]{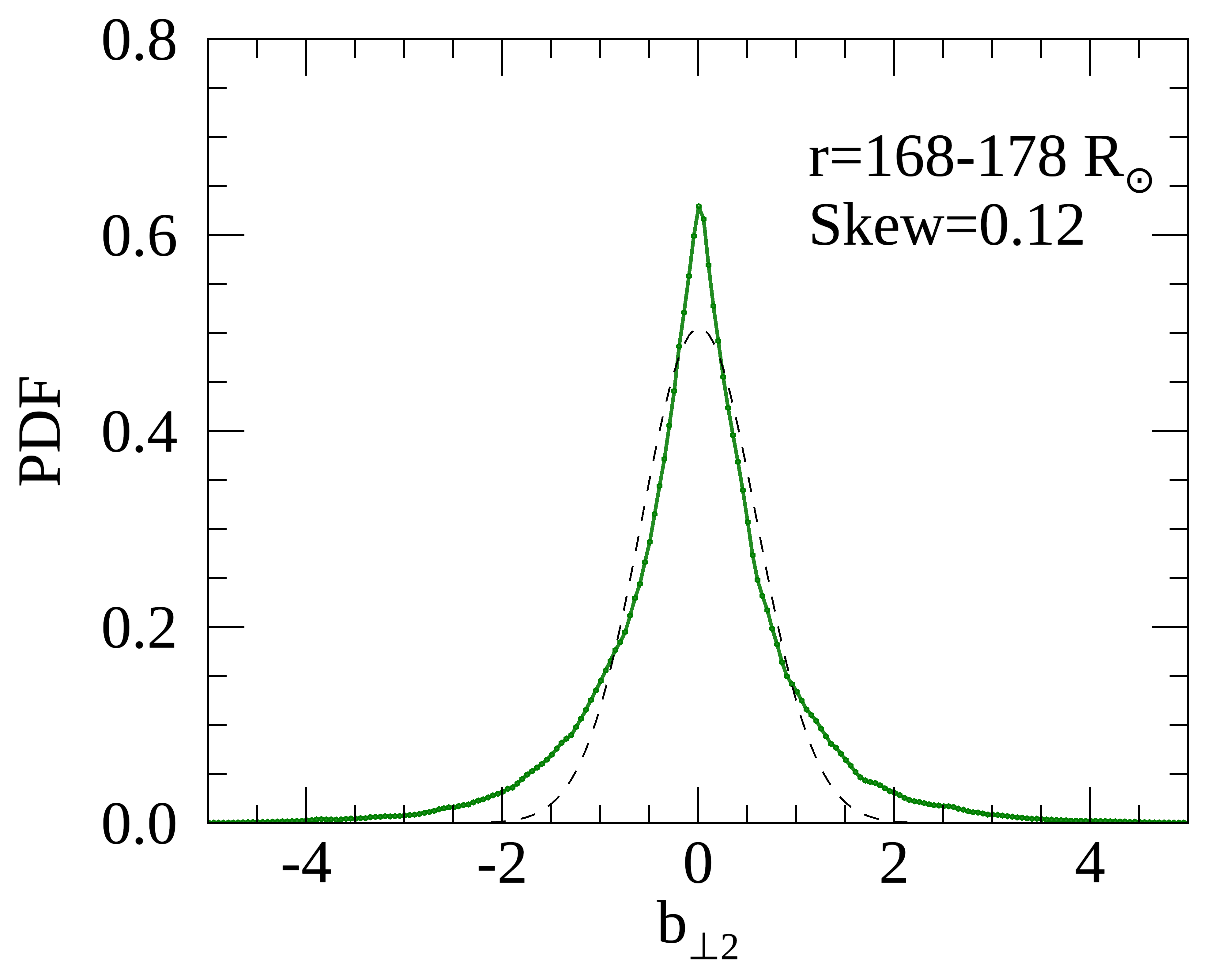}
    \caption{PDFs of magnetic fluctuations. Fields have been normalized by their respective standard deviations before computing PDFs. Bins with fewer than 10 counts are discarded. Dashed curve shows the best-fit Gaussian, which is fit to each PDF in the horizontal range \(\pm 3\) standard deviations. Skewness of each distribution is indicated. The chi-sqd goodness of fit to a Gaussian has values (in order of increasing \(r\)) 0.003, 0.002, 0.002 for \(b_\parallel\), and a value of 0.001 for \(\bperpa\) and \(\bperpb\), for all \(r\) shown.}
    \label{fig:pdf_b}
\end{figure*}

Finally, in the bottom panel of Figure \ref{fig:axisym} we test the commonly used assumption of axisymmetric turbulence in the plane transverse to the mean magnetic field, which would imply \(\delta \bperpa^2 / \delta \bperpb^2\approx 1\) \citep[e.g.,][]{matthaeus2007spectral,chhiber2021ApJ_flrw}. The PSP observations clearly indicate that this is not the case for all \(r\). In fact, a clear trend is observed, wherein \(\delta \bperpb^2\)  \(\sim 1.7\times \delta \bperpa^2\) at \(200~\rs\), and \(\delta \bperpa^2\) increases with decreasing \(r\), so that the ratio flips near \(50~\rs\), with \(\delta \bperpa^2 \sim 1.2 \times \delta \bperpb^2\) at \(30~\rs\). Recall that below \(\sim 60~\rs\) the mean field is nearly radial, so that \(\bperpa\) and \(\bperpb\) represent fluctuations in the \(T\) and \(N\) directions, respectively (see Section \ref{sec:mfc}). Therefore, the results indicate that transverse fluctuations in the ecliptic plane are slightly larger compared to the out-of-plane fluctuations, close to the Sun.\footnote{At larger helioradii the mean field does not lie in the ecliptic plane (see Figure \ref{fig:B0}), and no straightforward association can be made between the two transverse fluctuations and the ecliptic plane.} The value of \(\delta \bperpa^2/\delta\bperpb^2\) observed  by PSP near \(200~\rs\) is comparable to the value of \(0.87\) reported by \cite{padhye2001JGR} using Omnitape data. We note that the observed non-axisymmetry ratio ranges between \(0.6\dash 1.2\) for the distances considered, which implies that the two transverse variances are comparable, and the assumption of axisymmetry about the mean magnetic field may therefore be considered approximately valid in the inner heliosphere. Figure \ref{fig:axisym} also shows the ratio of the kurtoses of \(\bperpa\) and \(\bperpb\), which roughly stays close to unity for nearly all \(r\). 

The origin of the observed (modest) non-axisymmetry and its radial trend is unclear, although we briefly mention two possibilities here. First, the radial trend could indicate asymmetry between northern and southern solar hemispheres, since PSP's trajectory crosses from positive to negative heliolatitudes at a distance of about \(60~\rs\) in its first five orbits \citep[e.g.,][]{chhiber2021ApJ_psp}; these crossings roughly correspond to crossings of the Heliospheric Current Sheet (HCS) into regions of opposite magnetic polarity, since the HCS lies roughly at 0\degree~heliolatitude during solar minimum \citep[e.g.,][]{owens2013lrsp}. A similar north-south asymmetry has been observed in the winding angle of the interplanetary magnetic field spiral, which is more tightly wound north of the HCS  \citep{bieber1988JGR}. A second possibility is that, since \(\bperpa\) lies approximately in the \(\pm T\) direction below \(60~\rs\), strong azimuthal flows in this region \citep{kasper2019Nat} could produce large fluctuations in the magnetic field along the \(T\) direction. 
Note that non-axisymmetry of turbulence has implications for energetic particle transport \citep[e.g.,][]{ruffolo2008ApJ}.


%
\subsection{Radial evolution of PDFs of magnetic fluctuations}\label{sec:pdf}

We move on to an examination of the radial evolution of PDFs of the fluctuations. Figure \ref{fig:pdf_b} shows these PDFs at three selected ranges of helioradii -- \(28~\dash~38~\rs\), \(88~\dash~98~\rs\), and \(168~\dash~178~\rs\). Each component is normalized by its respective standard deviation within the respective radial bin before computing the PDF. Bins with fewer than 10 counts have been discarded. Best-fit Gaussians to each PDF are also shown; the Gaussians are computed using a non-linear least-squares fit to each PDF within \(\pm 3\) standard deviations. The distributions of \(b_\parallel\) in the top panels clearly indicate a finite (negative) skewness, which increases with decreasing \(r\), as was also seen in Figure \ref{fig:moments} (see discussion in Section \ref{sec:disc}). Both \(\bperpa\) and \(\bperpb\) have PDFs that are well approximated by a Gaussian, and do not show any change with \(r\).\footnote{Recall that the standard deviation of the fluctuations increases with decreasing \(r\) (Figure \ref{fig:moments}), but this is not seen in the plotted PDFs due to normalization of each quantity by its respective standard deviation within a radial bin.} The chi-squared goodness statistic, listed in the caption of Figure \ref{fig:pdf_b}, also indicates a higher degree of Gaussianity for the transverse fluctuations. These results are broadly consistent with those from \citep{padhye2001JGR}, who used Ulysses and Omnitape data.

\section{Conclusions and Discussion}\label{sec:disc}

We have used PSP observations accumulated over the first five orbits to examine PDFs of magnetic fluctuations in the inner heliosphere, and to trace their radial evolution between \(\sim 0.14\) and 0.9 AU. Expressing the magnetic field in a (local) mean-field coordinate system permits an investigation of component anisotropy. Effects of averaging-interval size are carefully considered. 

We find that PDFs of fluctuations transverse to the mean field are well approximated by a Gaussian, while parallel fluctuations have non-Gaussian values of skewness and kurtosis. This (negative) skewness increases in magnitude from \(\sim 1\) at \(200\ \rs\) to \(\sim 2.6\) at \(30\ \rs\), while the kurtosis ranges between 10 and 20, without a clear radial trend. The turbulence energy increases by more than an order of magnitude as one approaches the Sun, from \(\sim 3\ \text{nT}^2\) at \(200\ \rs\) to \(\sim 50\ \text{nT}^2\) at \(30\ \rs\). However, the ratio of the rms fluctuation to the mean magnetic field magnitude (``\(\delta b/B_0\)'') decreases from \(\sim 0.8\) to 0.5 between these distances, and is well described by a \(r^{1/4}\) power law. An increasing trend in variance anisotropy is observed with decreasing \(r\), with stronger fluctuations perpendicular to the mean field. A modest degree of  non-axisymmetry is observed in the plane transverse to the mean field, with a radial trend whose origin is unclear. 

Our results near 1 AU are broadly consistent with prior studies  \citep{marsch1994AnGeo,padhye2001JGR,bruno2003JGR}. The increasing skewness in the parallel fluctuation as one approaches the Sun indicates stronger non-linearities (which are related to third-order moments), and consequently a stronger turbulent cascade \citep[e.g.,][]{zhou2004RMP}. Indeed, \cite{bandyopadhyay2020ApJS_cascade} find that the energy transfer rate near PSP's first perihelion is about 100 times larger than the average value near Earth. \added{Further, since the parallel magnetic fluctuation is associated with compressible turbulence \citep[e.g.,][]{chen2012ApJ}, its increasing skewness raises the possibility of a stronger compressible cascade near the Sun \citep[see also][and references within]{brodiano2022arXiv}.} 

At the same time, our findings raise the question of the \textit{origin} of the observed skewness, \added{and whether this origin is solar or in-situ} -- if fluctuations at the solar source are initially Gaussian, then some in-situ process is required to generate skewness. \added{One possible explanation could lie in the increasing Alfv\'enicity of fluctuations observed by PSP as it approaches the Sun \citep{chen2020ApJS}. Alfv\'enicity implies  near constancy of \(|\bm{B}|\), and if there are large transverse fluctuations then the parallel fluctuation must be opposite to the mean field direction, thus giving rise to the observed skewness. This ``one-sided'' aspect of Alfv\'enic fluctuations was noted by \cite{gosling2009ApJ} at 1 au, and may be related to formation of magnetic switchbacks \citep[e.g.,][]{matteini2014GRL}, which also lead to a skewed distribution of the radial magnetic field \citep{ruffolo2020ApJ}.}


As PSP dives deeper into the corona, it will become possible to extend this type of study to distances within and below the trans-Alfv\'enic region \citep{kasper2021prl,chhiber2022MNRAS}. A preliminary study \citep{bandyopadhyay2022ApJ} has found that sub-Alfv\'enic plasma has stronger variance anisotropy compared to super-Alfv\'enic plasma. One may expect \(\delta b/B_0\) to keep decreasing deeper within the low-\(\beta\) corona as well \citep{chhiber2019psp2}. Some models of switchback generation predict that the occurrence of switchbacks will decrease below the trans-Alfv\'enic region \citep{ruffolo2020ApJ,schwadron2021ApJ,pecora2022ApJ}, and it will be interesting to examine whether this could be reflected in the skewness of the parallel fluctuation. Finally, later PSP orbits during solar maximum are expected to sample more data intervals with fast wind, which would provide  opportunities to compare the properties of PDFs of fluctuations in slow and fast wind \citep{padhye2001JGR}.


\begin{acknowledgments}
The author thanks T. N. Parashar, L. Matteini, and W. H. Matthaeus for useful discussions, and acknowledges the Parker Solar Probe (PSP) mission for use of the data, which are publicly available at the \href{https://spdf.gsfc.nasa.gov/}{NASA Space Physics Data Facility}. This research is partially supported by NASA under the Heliospheric Supporting Research program grants 80NSSC18K1210, 80NSSC18K1648, and 80NSSC22K1020, and by the PSP Guest Investigator program grant 80NSSC21K1765. 
\end{acknowledgments}



\end{document}